\documentclass{PoS}

\title{Searches for Ultra-High Energy Neutrinos\\with ANITA}

\ShortTitle{Searches for Ultra-High Energy Neutrinos with ANITA}

\author{\speaker{Cosmin Deaconu} for the ANITA collaboration\footnote{for collaboration list, see PoS(ICRC2019)1177 or  \tt{https://anitaneutrino.github.io/authorlist}
}\\
        Department of Physics, Enrico Fermi Institute, Kavli Institute for Cosmological Physics, University of Chicago\\
        E-mail: \email{cozzyd@kicp.uchicago.edu}}

\abstract{

The ANtarctic Impulsive Transient Antenna (ANITA) long-duration balloon
experiment flies an interferometric radio array over Antarctica with a
primary goal of detecting impulsive Askaryan radio emission from ultra-high-energy neutrinos interacting in the ice sheet. The third and
fourth ANITA flights were completed in January 2015 and December 2016,
respectively, obtaining the most stringent limits on the diffuse ultra-high-energy neutrino flux above 10$^{19.5}$ eV to date. We also discuss ongoing analyses and the proposed Payload for Ultrahigh Energy Observations (PUEO), the successor to the ANITA program. PUEO's larger number of antennas and improved trigger would significantly improve sensitivity compared to ANITA-IV. 
}

\FullConference{36th International Cosmic Ray Conference -ICRC2019-\\
		July 24th - August 1st, 2019\\
		Madison, WI, U.S.A.}

\begin{document}

\section{Introduction}

Ultra-high-energy (UHE) neutrinos ($>1$ EeV) may be produced by photonuclear processes involving the UHE high-energy cosmic rays and photons, either at the sources or during propagation with the cosmic microwave background~\cite{GZK1,GZK2}.
The low expected flux~\cite{kotera} and small cross-section~\cite{crosssection} imply that vast detection volumes are needed for a successful detection. The Askaryan mechanism~\cite{askaryan}, a coherent enhancement of the Cerenkov emission at long wavelengths resulting from negative charge excesses of particle showers in dense media, along with the long ($\mathcal{O}$(1 km)) radio attenuation length in ice,  allow the use of the abundant Antarctic ice sheet as a detection target for UHE neutrinos. 

The ANtarctic Impulsive Transient Antenna (ANITA) collaboration has been searching for the impulsive Askaryan emission from neutrinos interacting in the Antarctic ice sheet for more than a decade using a sensitive radio receiver deployed on a NASA long-duration ballooon payload. ANITA is also sensitive to the radio emission from extensive air showers (EAS)~\cite{anita1CR}, which is primarily due to the charge separation by the Earth's geomagnetic field. Such EAS may be produced from UHE cosmic-ray showers, although ANITA has also unexpectedly detected emission that may be consistent with upward-going EAS~\cite{mysteryEvent}.

\section{The ANITA Instrument}

\begin{figure}[b]

\begin{minipage}{0.27\columnwidth}
\centering
(a) ANITA-IV Payload\\ 
\includegraphics[width=1.9in]{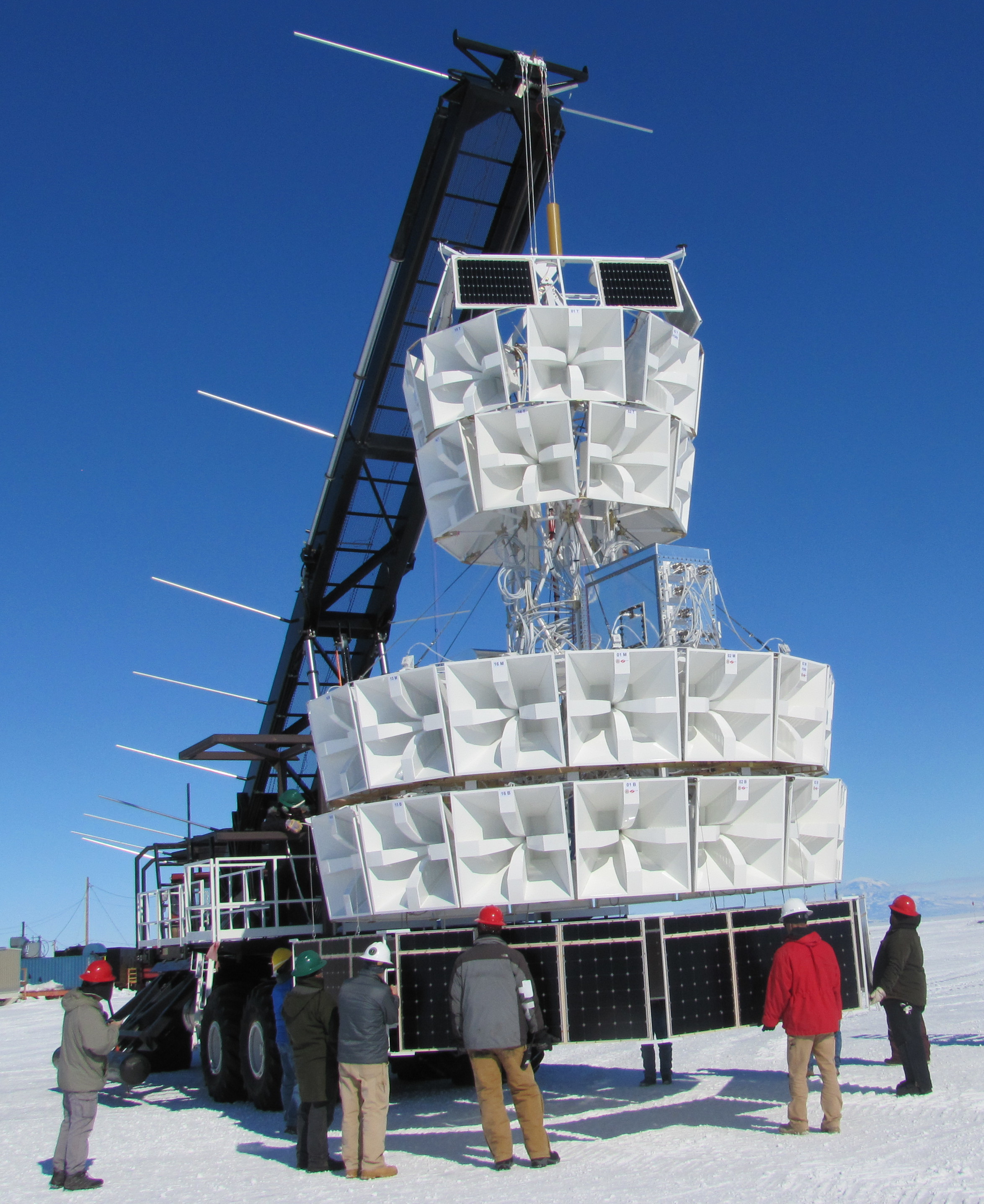}
\end{minipage}
\hfill
\begin{minipage}{0.7\columnwidth}
\centering
(b) ANITA-III Trigger\\
\includegraphics[width=3.7in]{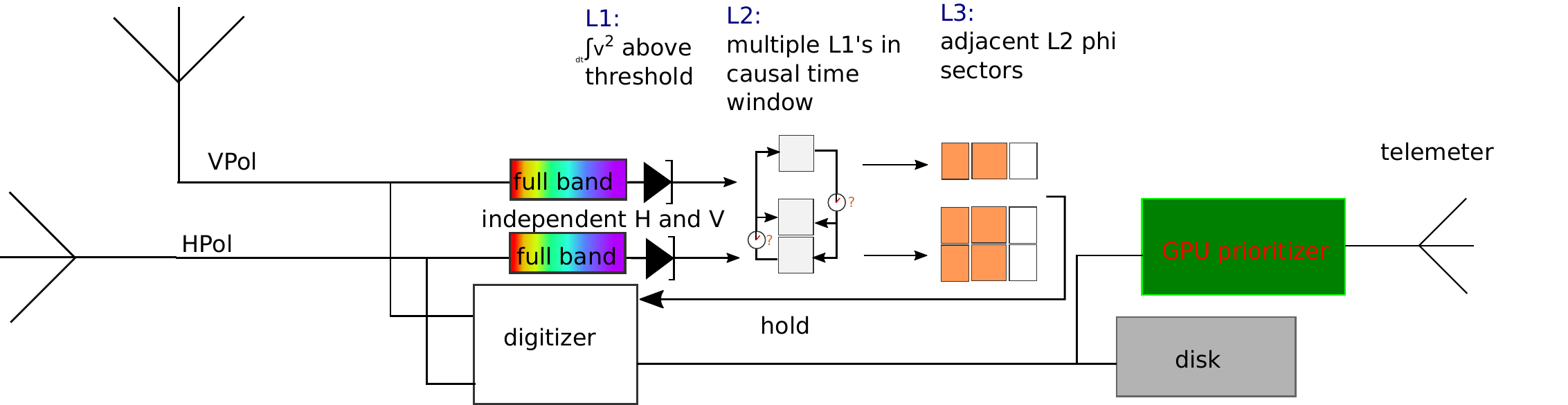}
\\
\vspace{0.1in}
(c) ANITA-IV Trigger\\
\includegraphics[width=3.65in]{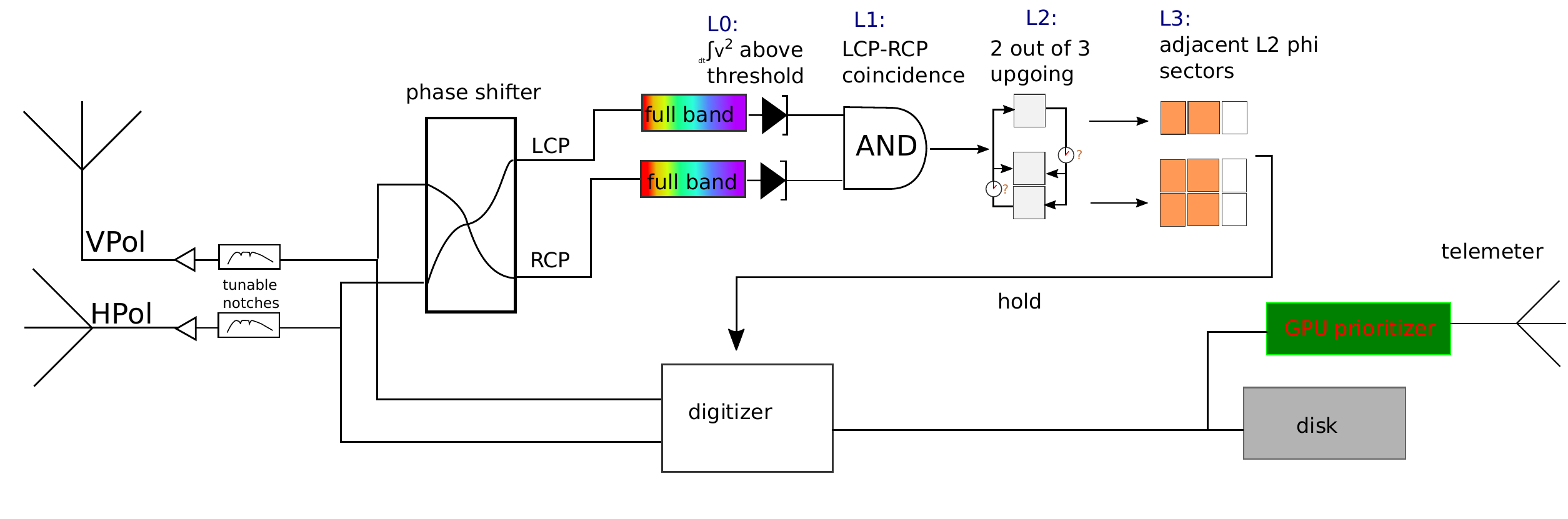}
\end{minipage}
\caption{(a): The ANITA-IV payload, prior to launch from LDB. ANITA-III has a similar mechanical design.  (b) and (c): schematic depictions of the trigger strategies for the ANITA-III and ANITA-IV. See text. }
\label{fig:anita} 
\end{figure}

ANITA has flown a total of four missions, all starting from the long-duration ballooning (LDB) facility on the Ross Ice Shelf near McMurdo Base: ANITA-I for 35 days in 2006-2007, ANITA-II for 30 days in 2008-2009, ANITA-III for 22 days in 2014-2015, and ANITA-IV for 29 days in 2016. We focus on the latter two flights here, which published diffuse neutrino search results~\cite{anita3,anita4} within the last year and a half. 

ANITA-III and ANITA-IV share similar mechanical designs (Fig.~\ref{fig:anita}(a)), each with 48 dual-polarization horn antennas arranged in an circular pattern with 16 azimuthal sectors, each with a top, middle, and bottom antenna. The antenna band is 180-1200 MHz. Both payloads use the LABRADOR~\cite{lab3} switched-capacitor array digitizers capable of sampling 260 samples at 2.6 GSa/s, to record the 96 input channels. ANITA rotates freely and uses a pair of differential GPS receivers to record the orientation during flight. Data is stored on a redundant hard drive array, with a small fraction of data telemetered during flight via satellite or a line-of-sight system. 
ANITA-III's nominal float altitude was 37 km while ANITA-IV had a lighter design and reached 40 km.  ANITA-III also incorporated a drop-down low-frequency antenna (30-80 MHz). 

ANITA-III and ANITA-IV primarily differ in the front-end and trigger designs, schematically depicted in Fig.~\ref{fig:anita}(b,c). For ANITA-III, after bandpass filtering and a low-noise amplifier (LNA) at each antenna port, the horizontal-polarization (Hpol) and vertical-polarization (Vpol) triggers operate independently, each capable of triggering readout. The channel-level trigger (L1) compares the output level of a tunnel diode used as a square-law detector with a dynamically-set threshold, where the threshold is set by a control loop to maintain an L1 rate of 450 kHz on each channel. The second-level trigger (L2) requires at least two of the three antennas of the same polarization in each azimuthal sector to have an L1 trigger within a causal time window for upgoing events. Finally, the global (L3) trigger requires two same-polarization L2 triggers within 10 ns. The ideal trigger rate is as high as possible without incurring significant deadtime, $\mathcal{O}$(50 Hz).

ANITA-III suffered from significant continuous wave (CW) contamination from geostationary military communications satellites launched since the ANITA-II flight. To prevent individual azimuthal sectors from monopolizing the trigger and inducing excessive deadtime, the ANITA flight software can automatically mask azimuthal sectors with too high a trigger rate. Due to these satellites, much of the north-facing side of the payload was effectively constantly masked. 

To mitigate CW contamination, a set of Tunable Universal Filter Frontends (TUFFs)~\cite{TUFF} were developed for ANITA-IV. Each TUFF consists of a set of three notch filters in relevant ranges for CW that may be individually tuned in flight by adjusting variable capacitors. In addition to the TUFFs, ANITA-IV enhancements include lower noise figure due to an improved LNA/frontend design, and a modified trigger strategy more sensitive to signals that are not purely Hpol or Vpol.  Rather than triggering directly on Hpol or Vpol, ANITA-IV employs 45$^\circ$ phase shifters in the trigger path to convert each antenna's signal to left-and-right circular polarization (LCP, RCP). The primarily linearly-polarized signals expected from neutrinos or EAS should have approximately equal LCP and RCP, while incoherent thermal noise or circularly-polarized CW from satellites in general will not. The channel-level trigger (now L0) maintains the tunnel diode square-law detector but an additional step (L1) is added, requiring L0s from both LCP and RCP within 4 ns. This allows ANITA to decrease the effective threshold while being more sensitive to mixed-polarization signals. The L0 rate for ANITA-IV is maintained at roughly 6 MHz. The L2 and L3 trigger requirements are virtually the same as ANITA-III, with some adjustments to the L2 time windows.  
\section{Calibration and Simulation}

\begin{figure}
\centering
\includegraphics[height=1.5in]{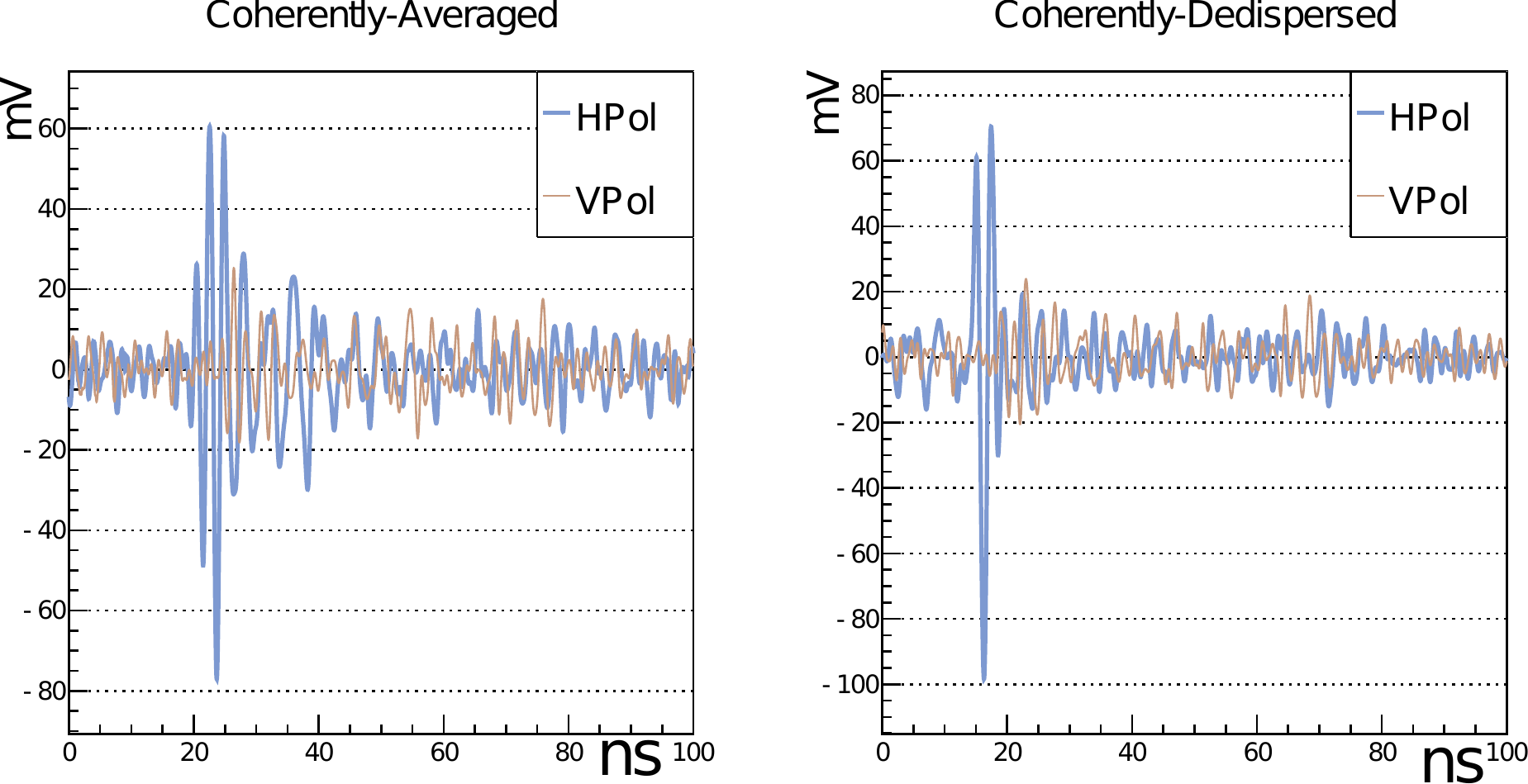}
\hspace{0.1in}
\vline
\hspace{0.1in}
\includegraphics[height=1.5in]{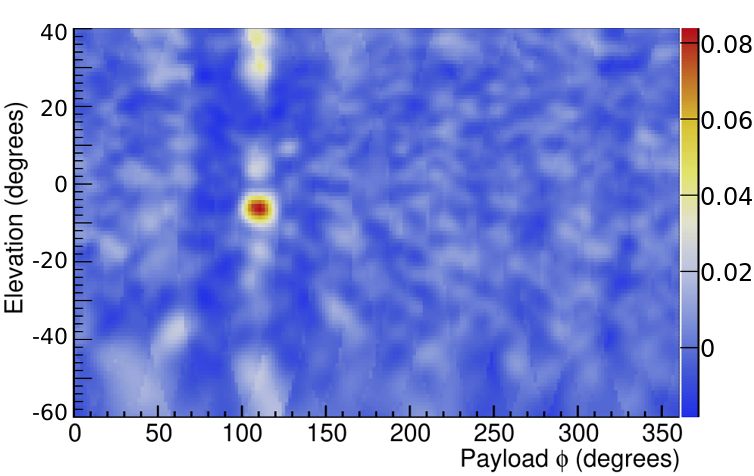}

\caption{(Left) Signals from a calibration pulser deployed at WAIS divide during the ANITA-III flight. (Right) The interferometric reconstruction map associated with this event. The color axis indicates the average correlation between antennas with time delays applied for each direction.} 
\label{fig:event}
\end{figure}

In addition to lab measurement of channel responses and digitizer properties and high-voltage pulsing just after launch, high-voltage pulsers are deployed on the Antarctic continent along the expected flight path in order to calibrate the detector. Pulses are synchronized to the GPS second to simplify identification. Additionally, the companion HiCal~\cite{hical,hical2_instrument} trailing payload produces high-voltage pulses that can trigger the payload, both directly or by first reflecting off the ice.  In addition to RF triggers, both payloads record forced triggers at 3 Hz which are useful for understanding the unbiased RF environment.  

ANITA's acceptance to UHE for each flight is estimated by \texttt{icemc}~\cite{icemc} and another independent simulation. The simulations model interactions of neutrinos in the ice, the radio emission from interaction, propagation to the payload and the trigger logic for each payload (and its particular path). The digitizer path is also modeled so that analysis efficiencies may be estimated.

\section{Analysis Methods}
Both ANITA-III and ANITA-IV produced $\mathcal{O}(10^7)$ total RF triggers. Most of these are due to either thermal noise or anthropogenic signals from satellites, the payload itself, or human activity in Antarctica. At most a few of the signals are likely to be of neutrino origin and perhaps several dozen of EAS origin, therefore many orders of magnitude of data reduction must be performed. ANITA-III and IV both had multiple independent, blind neutrino analyses performed by different teams. Full details of the completed diffuse flux analyses are available in ~\cite{anita3} and ~\cite{anita4}, but some of the basic procedure common to all analyses is summarized here. All analyses made use of the calibration data and simulated datasets to optimize the searches.

After removing digitizer glitches and events triggered by RFI from the payload (identifiable by the amplitude pattern across antennas), each analysis employed adaptive CW filtering. Even if CW is not responsible for triggering an event, it may still be present and confound further analysis, so frequencies with excessive power are removed with special care taken not to distort the waveform.  

Next, each analysis constructs observables used to distinguish between impulsive broadband events typical of Askaryan neutrinos and other classes of events (thermal noise, pure CW, non-impulsive anthropogenics). Many of these observables are derived from an interferometric map ~\cite{interferometric}(see Fig.~\ref{fig:event}, right), where for all possible directions, the average correlation of all antenna pairs is computed using the appropriate time delays for each baseline for a plane wave from that direction. The peak of the map is interpreted as the likely source of the emission generating the event, and a coherent sum is formed in that direction, from which additional observables are derived. Calibration data indicates pointing resolutions down to 0.1 degrees in elevation and 0.3 degrees in azimuth for large signals. One, or more generally, a multivariate combination of observables are used at this stage to reduce the dataset to only impulsive, broadband events. 

Many anthropogenic signals from human activity on the continent are also broadband and impulsive. Unlike the rare physics events, anthropogenic events tend to cluster spatially on the continent. Each analysis therefore either excludes regions of ice with multiple originating events or sets cuts differently for each patch of Antarctica. Only events that are extreme outliers for their neighborhood will survive final cuts. 

Finally, the Askaryan neutrino candidates must be separated from the radio emission from air showers. As the geomagnetic field in Antarctica is primarily vertical, emission from EAS is primarily Hpol. Conversely, due to absorption of UHE neutrinos in the Earth and the emission geometry, radio emission from Askaryan neutrinos is primarily Vpol. Only primarily Vpol events are considered candidates for signals from Askaryan neutrinos.  The background estimate and analysis efficiency are determined prior to unblinding.

EAS candidates are required to be isolated, impulsive and predominantly Hpol. As EAS have been detected before and the basic waveform properties are known, we additionally require that the waveform shape be consistent with expectation. In ANITA-IV, we additionally constrain the polarization angle to lie near the geomagnetic expectation. For cosmic-ray induced EAS, which occur at altitudes well below ANITA, two geometries are possible: ``direct" earth-skimming EAS and ``reflected" EAS where the radio signal is reflected off the ice. The latter is more common, and a polarity flip is expected between the two geometries. While cosmic-ray-induced EAS are mundane at this point, due to the potential physics implications of detecting upgoing EAS the polarity of events is randomized during analysis as a form of blinding. 

\section{Results}
\begin{figure}[t] 
\begin{minipage}{0.4\columnwidth}
\centering
(a) ANITA-III candidate\\
\includegraphics[trim=4.4in 0in 0in 0.2in,clip,width=2in]{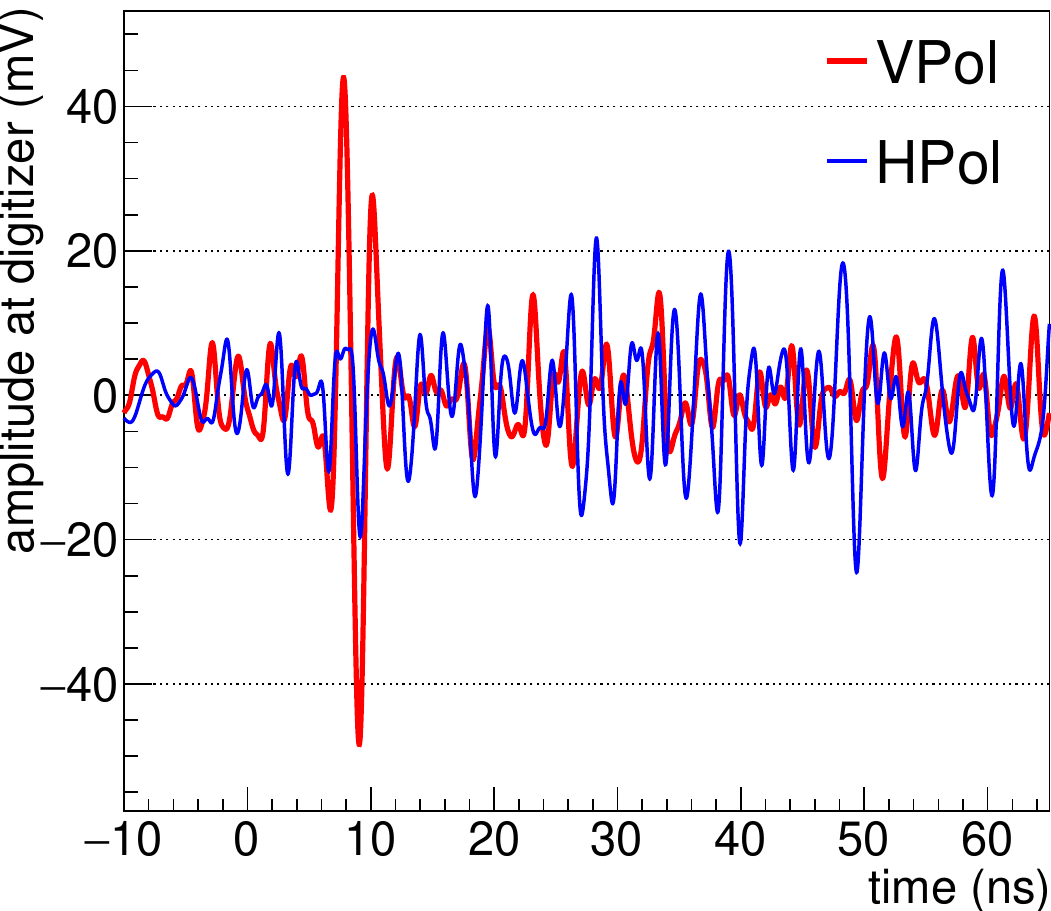}\\
(b) ANITA-IV candidate
\includegraphics[trim=0in 4.8in 0in 0.45in,clip,width=2.5in]{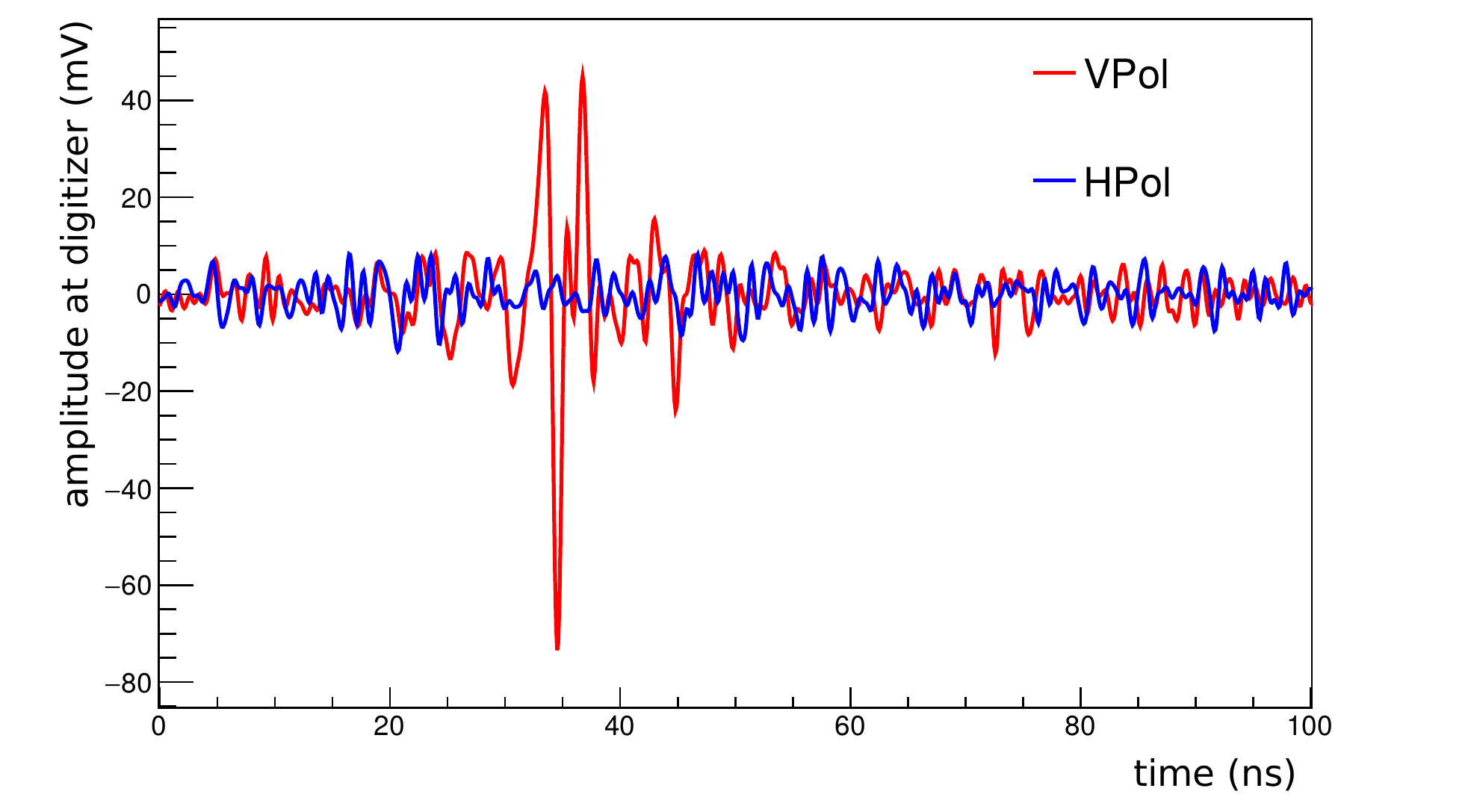}
\end{minipage} 
\hfill 
\begin{minipage}{0.6\columnwidth}
\centering
(c)\\ 
\includegraphics[width=3.5in]{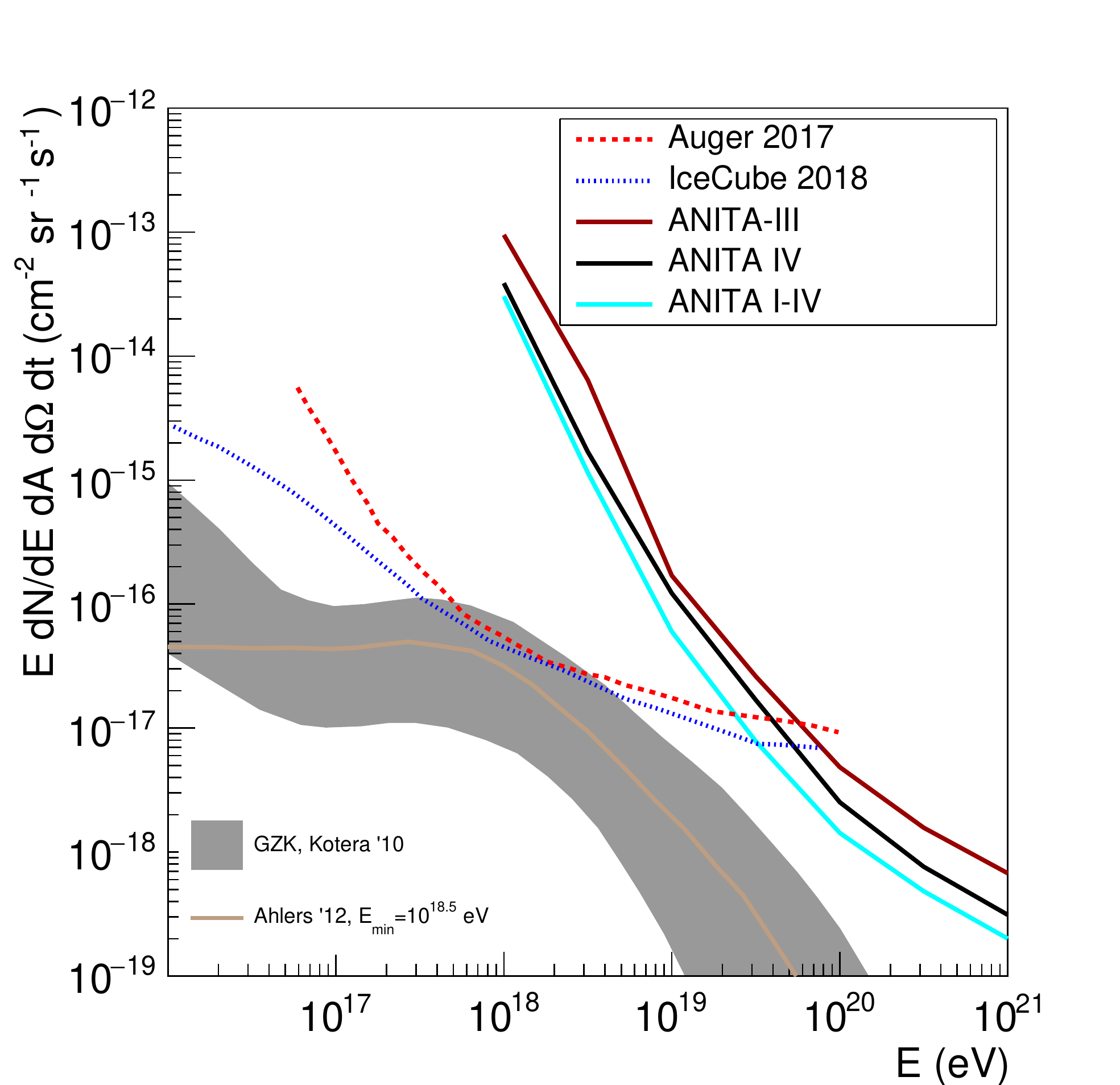}
\end{minipage} 

\caption{(a) The event passing all cuts for the most sensitive ANITA-III diffuse neutrino analysis. (b) Same for ANITA-IV. Both events are impulsive, primarily VPol and isolated, but nonetheless consistent with the background expectation from each flight. (c) The limit plots for ANITA-III, ANITA-IV and the combination of ANITA I-IV along with some cosmogenic models and limits from Auger and IceCube~\cite{kotera}~\cite{ahlers}~\cite{icecube2017erratum}~\cite{auger2017}}
\label{fig:results} 

\end{figure}

The most sensitive of the ANITA-III diffuse neutrino searches, based on the pre-unblinding expected sensitivity, found one candidate Vpol event on an expected background of 0.7$^{+0.5}_{-0.3}$ (Fig.~\ref{fig:results}a), with alternative analyses producing results consistent with expected background estimates and efficiencies. 
The most sensitive of the ANITA-IV diffuse neutrino searches also found one candidate event on an expected background of 0.64$^{+0.69}_{-0.45}$ (Fig.~\ref{fig:results}b). An alternative analysis found a different candidate event, but this is not inconsistent with the mutual efficiency and background estimate of each analysis. In both cases, the primary contribution to background is from the possibility of isolated, impulsive anthropogenic events. Due to the lower threshold, reduced susceptibility to CW, and longer flight time, ANITA-IV had a significantly improved expected exposure to UHE neutrinos. By interpreting the candidate events as background, an upper limit may be set on the flux of UHE neutrinos. Fig.~\ref{fig:results}c shows the limits from ANITA-III, ANITA-IV, and all flights combined. 

Both ANITA-III and ANITA-IV detected 20-30 EAS candidates. While ANITA-IV's increased sensitivity and flight time was expected to garner many additional EAS candidates, the use of the tunable notch filters led to some loss of sensitivity to EAS, which are more low-frequency weighted than signals expected from neutrinos. After polarity unblinding, one ANITA-III EAS candidate had inverted polarity with respect to its geometric expectation~\cite{anita3me}, on a predicted background for the EAS analysis of $<10^{-2}$ . Like the ANITA-I candidate, it was steeply upward going (35 degree below the horizontal) and very far ($\mathcal{O}(100\sigma)$)  from the horizon. While such an upward-going EAS could be induced by the decay of a $\tau$ from a $\tau$ neutrino, given the long chords through the earth and tension with other experiments, this hypothesis is disfavored for a diffuse flux~\cite{anita_tau,steph_icrc}. 

\section{Ongoing Searches}

While the set of the ANITA-IV EAS have been identified, the unblinded polarity analysis is not yet complete. The primary reason for this is a calibration issue. Unlike previous flights, the use of the tunable notch filters means that the system response of ANITA-IV changed during flight and a polarity comparison between EAS events requires deconvolution of the system response. As the notch filter configurations could not be known prior to flight, this calibration had to be done afterwards. Moreover, the improvements in noise figure of the frontend amplifiers led to a more complicated interaction between the antenna response and the rest of the system response than in previous flights, the characterization of which is still in progress. 

Additionally, searches for neutrinos coincident with multimessenger observations are in progress on the ANITA-III and ANITA-IV datasets. By focusing the searches to times and directions of transient astrophysical phenomena that may potentially produce UHE neutrinos, the analysis efficiency can be increased (particularly at lower energies). Sources considered include GRBs, flaring blazars, and putative sources identified by IceCube~\cite{icecube_mm}. Intriguingly, the potential 2014 neutrino flare reported by IceCube~\cite{icecube_flare} occurred during the ANITA-III flight. Preliminary simulation studies suggest that a regression using the polarization vector can be used to reconstruct neutrino direction to several degree resolution in both RA and declination.  We expected these analyses to be completed in the coming months.

\section{Future: Payload for Ultra-high Energy Observations (PUEO)}
\begin{figure}
\centering
\includegraphics[height=2.8in]{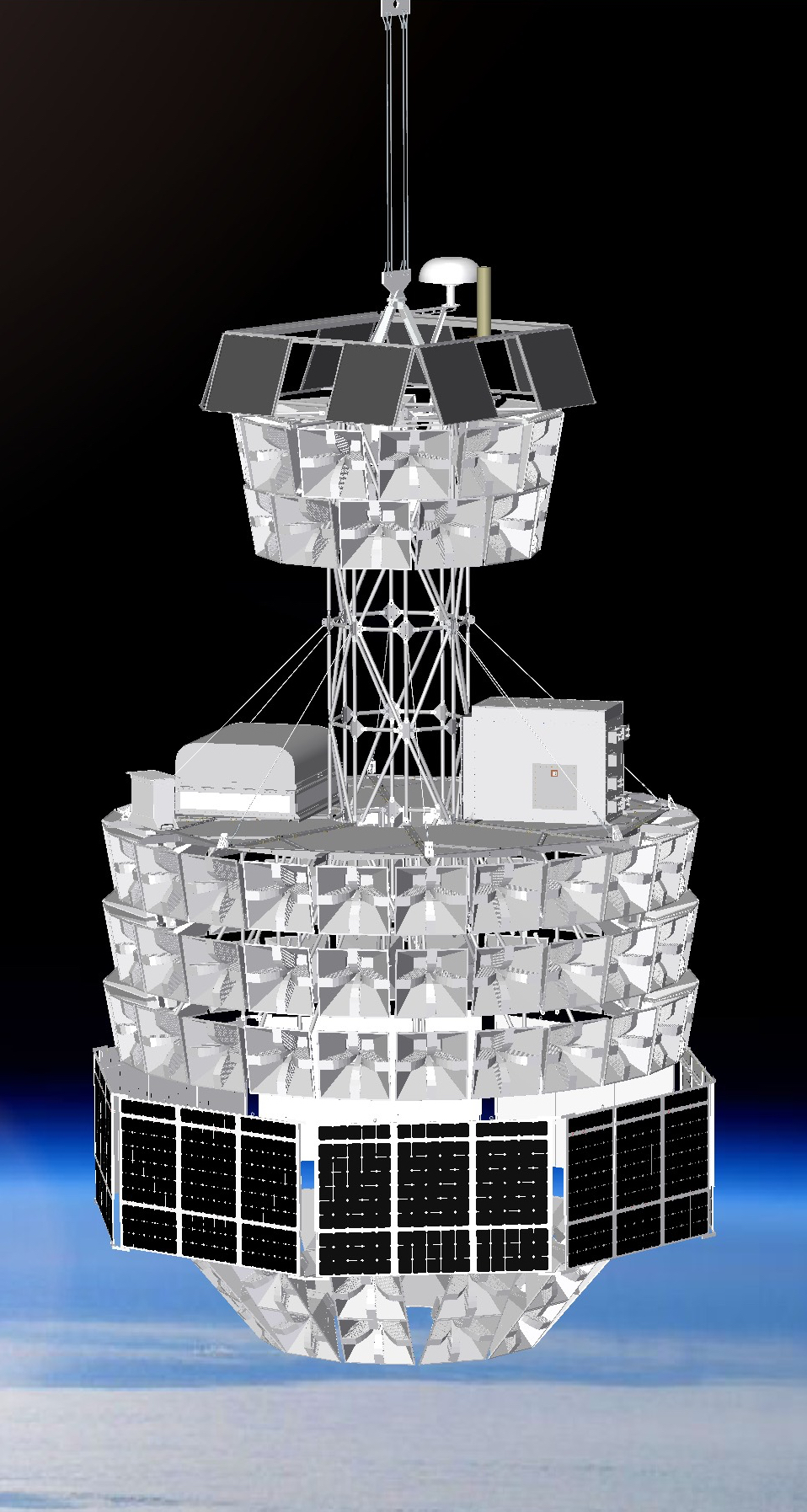}
\hspace{1cm}
\includegraphics[height=2.8in]{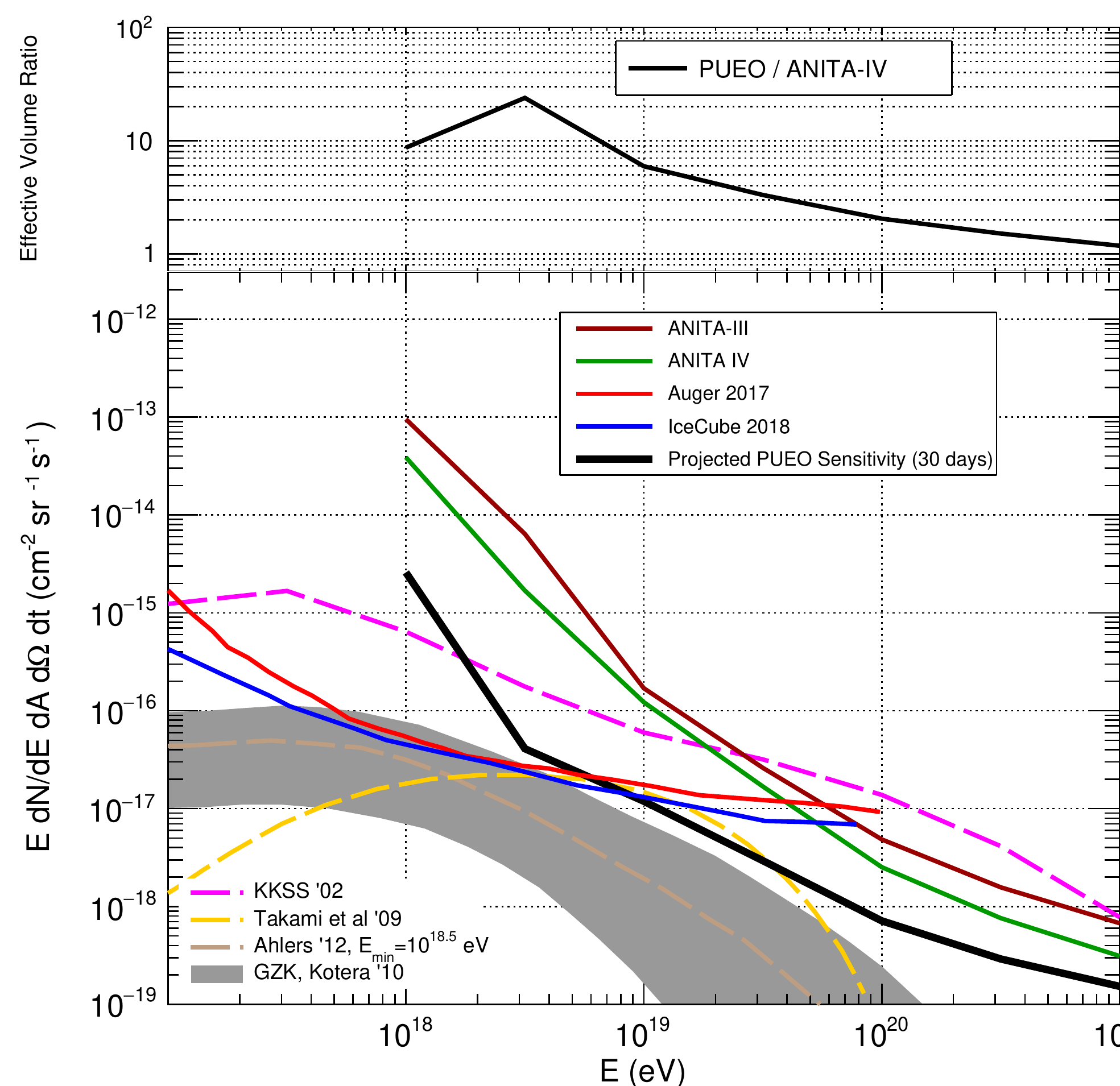}
\caption{(Left) A rendering of the proposed PUEO design. (Right) The projected Askaryan neutrino sensitivity of PUEO.} 
\label{fig:pueo} 
\end{figure} 

The proposed Payload for Ultra-high Energy Observations (PUEO) is the next generation of the ANITA concept. PUEO 's design incorporates 120 dual-polarization antennas, although with a higher cutoff frequency (300 MHz) so that more antennas can fit within the launch envelope. A rendering of the PUEO payload is shown in Fig.~\ref{fig:pueo} (left). 

Rather than the square-law detector combinatoric trigger previously used by ANITA, PUEO will employ a beamforming trigger. In the trigger path, signals from 16 antenna clusters will be electronically summed with delays corresponding to sets of trial directions (beams) covering all directions. This phased array technique has been previously investigated by ANITA~\cite{andres_proceedings} and successfully been demonstrated by ARA to reduce the trigger threshold~\cite{phasedarray}. To fit within the power budget available, low-bit streaming digitizers will be used in the trigger path. Simulations indicate that PUEO's trigger threshold, in peak-to-peak relative to thermal noise, will be a factor of 2.5 lower than ANITA-IV.  This manifests itself in greatly improved sensitivity, especially in the lower energy range, as shown on the right of Fig.~\ref{fig:pueo}.  

While the low-bit digitizers are sufficient for triggering purposes, they lack the dynamic range needed for digitization. The improved LAB4d~\cite{lab4} chips, with much improved high-frequency response, will be used in the digitizer path. A version of the TUFFs will still be used, although only on the trigger path.  

Ninety-six of PUEO's antennas are part of the main interferometric trigger. PUEO also contains 24 antennas canted more steeply downwards at 40$^\circ$, in order to fill a gap in elevation coverage. These antennas would have a separate, low-rate trigger similar to ANITA's trigger and are primarily for reconstruction and EAS detection.

\section*{Acknowledgements}

ANITA is funded by NASA grant NNX15AC24G and relies on the NSF for logistics support. We particularly thank the team at the Columbia Scientific Ballooning Facility (CSBF) responsible for keeping ANITA aloft and the NSF personnel in Antarctica who are dedicated to making science happen on a harsh continent. 
 This work  was  supported  by  the  Kavli  Institute  for  Cosmological  Physics  at  the  University  of  Chicago.   Computing  resources  were  provided  by  the  Research  Computing  Center  at  the  University  of  Chicago.

\bibliographystyle{custom_style}
\bibliography{main}

\end{document}